\documentclass[twocolumn,superscriptaddress,showpacs,floatfix,preprintnumbers, nofootinbib,hyperref]{revtex4} 
\usepackage{graphicx}
\usepackage{rotating}
\usepackage{epsfig}
\usepackage{bm}
\usepackage[usenames,dvipsnames]{xcolor}

\usepackage{color}

\def\he4{$^4$He}
\def\h2{$^2$H}

\newcommand{\lesssim}{\,\rlap{\lower3.7pt\hbox{$\mathchar\sim$}}
\raise1pt\hbox{$<$}\,}

\begin{document}

\preprint{IPPP/14/24, DCPT/14/48}

\title{Damping the neutrino flavor pendulum by breaking homogeneity}

\author{Gianpiero Mangano}
\affiliation{Istituto Nazionale di Fisica Nucleare - Sezione di Napoli,
Complesso Universitario di Monte S. Angelo, I-80126 Napoli, Italy} 

\author{Alessandro Mirizzi} 
\affiliation{II Institut f\"ur Theoretische Physik, Universit\"at Hamburg, Luruper Chaussee 149, 22761 Hamburg, Germany}

\author{Ninetta Saviano} 
\affiliation{Institute for Particle Physics Phenomenology, Department of Physics,
Durham University,\\ Durham DH1 3LE, United Kingdom}


\begin{abstract}

The most general case of self-induced neutrino flavor evolution is described by a set of kinetic
equations for   a dense neutrino gas  
evolving both in space and time. 
 Solutions of these equations have been typically worked out assuming that either 
the time (in the core-collapse supernova environment) or space (in the early universe) homogeneity in the initial conditions is preserved through the evolution. In these cases 
one can gauge away the homogeneous variable and reduce the dimensionality of the problem.  In this paper
we investigate if small deviations from an initial postulated homogeneity can be amplified by the interacting
neutrino gas,  leading to a new flavor instability. To this end, we consider a simple two flavor isotropic neutrino gas  evolving in time, and initially composed by only  $\nu_e$ and
$\bar\nu_e$ with equal densities. In the  homogeneous case,  this system  shows a bimodal instability in the inverted mass hierarchy scheme, leading to the well studied flavor pendulum behavior.
This would lead to periodic pair conversions $\nu_e \bar\nu_e \leftrightarrow \nu_x \bar\nu_x$. 
 To break  space homogeneity, we  introduce small amplitude space-dependent perturbations in the matter potential.
By Fourier transforming the equations of motion with respect to the space coordinate, we then numerically solve  a set of coupled equations for the different Fourier modes. 
We find that even for arbitrarily tiny inhomogeneities,   the system evolution runs away from the stable pendulum behavior: the different modes are excited  and the space-averaged ensemble evolves 
 towards flavor equilibrium. We finally comment on the role of a time decaying neutrino background density in weakening these results.
 \end{abstract}


\maketitle

\section{Introduction}

Neutrino--neutrino interactions in dense neutrino media are known to produce surprising flavor oscillation effects,
in the form of self-induced conversions, when the typical neutrino self-interaction potential
$\mu = {\sqrt{2} G_F}n_\nu$ is comparable with or greater than the vacuum oscillation frequency $\omega=\Delta m^2/2E$ (see e.g.~\cite{Duan:2010bg} for a recent review). 
This situation can be encountered in the early universe or in core-collapse supernovae (SN), where neutrino
themselves form a background medium for their propagation.
Differently from the usual Mykheyeev-Smirnov-Wolfenstein (MSW) effect~\cite{Matt}, associated with the
matter potential $\lambda=\sqrt{2}G_F n_e$, the self-induced effects  do not change the flavor content of the neutrino ensemble.
Yet, the flavor is exchanged between different momentum modes, leading to peculiar spectral features
known as  {\it spectral swap} and {\it split}~\cite{Raffelt:2007cb}. 

The growth of these  effects is associated with instabilities in the flavor space, which are amplified
by the neutrino-neutrino interactions~\cite{Sawyer:2005jk,Banerjee:2011fj}.
An example is represented by the bimodal instability~\cite{Samuel:1995ri} of an isotropic and homogeneous  dense  gas  of neutrinos and antineutrinos in equal
amounts. 
They convert  from one flavor to another in  pair production processes
$\nu_e \bar\nu_e \leftrightarrow \nu_x \bar\nu_x$,  behaving as a flavor pendulum 
even if the mixing angle is very small~\cite{Hannestad:2006nj,Duan:2007fw}. In this case,  the vacuum mixing angle acts as a seed triggering the flavor instability. 
In non-isotropic neutrino gases, like the case of neutrinos streaming-off a SN core,  the features of the self-induced effects are more involved, since
the current-current nature of the low-energy weak interactions introduces
an angle dependent term  $(1-{\bf v}_{\bf p} \cdot {\bf v}_{\bf q})$, where ${\bf v}_{\bf p}$ and  ${\bf v}_{\bf q}$
are neutrino velocities~\cite{Qian:1994wh,Duan:2006an}.
It has been shown that this term  can lead to a  {\it multi-angle} instability, 
which hinder the maintenance of the coherent oscillation 
behavior for different neutrino modes~\cite{Duan:2006an,EstebanPretel:2007ec,Sawyer:2008zs}. In particular, in a symmetric gas of equal neutrino and antineutrino densities
even a very small anisotropy is sufficient to trigger a run-away towards flavor equipartition~\cite{decoh}.
An additional instability has been recently discovered in the SN context. Removing the assumption of axial symmetry in the $\nu$ propagation, a  multi-azimuthal-angle instability
emerges, even  assuming a perfect spherically symmetric $\nu$ 
emission~\cite{Raffelt:2013rqa,Raffelt:2013isa,Duan:2013kba,Mirizzi:2013rla,Mirizzi:2013wda,Chakraborty:2014nma}.  

Symmetries in the neutrino self-induced evolution are often assumed in order to  reduce the complexity of the problem.
Nevertheless,  these recent findings {\it question} the validity of these assumptions, since they suggest that (unavoidable) small deviations from initial  symmetries could be dramatically amplified
by the interacting neutrinos during the evolution. 
In absence of collisions, the dynamics
of the  $\nu$ space-dependent occupation numbers or Wigner function $\varrho_{{\bf p}, {\bf x}}(t)$
with momentum ${\bf p}$ at position ${\bf x}$ is ruled by the kinetic equations~\cite{Sigl:1992fn,Strack:2005ux} 

\begin{eqnarray}
&& \partial_t \varrho_{{\bf p}, {\bf x}} + {\bf v}_{\bf p} \cdot \nabla_{\bf x} \varrho_{{\bf p}, {\bf x}} 
+ {\dot{\bf p}} \cdot \nabla_{\bf p} \varrho_{{\bf p}, {\bf x}}    \nonumber \\
&& = - i [\Omega_{{\bf p}, {\bf x}}, \varrho_{{\bf p}, {\bf x}}]
\,\ ,
\label{eq:eom}
\end{eqnarray}
with the Liouville operator in the left-hand side. In particular, the first term accounts for an explicit time
dependence, while the second is the drift term proportional to the neutrino velocity
${\bf v_p}$, due to particle free streaming. Finally, the third term is  proportional to the force acting on neutrinos. 
On the right-hand-side the matrix $\Omega_{{\bf p}, {\bf x}}$ is the full Hamiltonian containing the
vacuum, matter and self-interaction terms. We remind the reader that the quantum-mechanical uncertainty between location and momentum
implies that this formalism can be applied only for cases  where spatial variations of the ensemble 
are weak on the microscopic length-scale defined by the typical particle wave-length.

In general, Eq.~(\ref{eq:eom}) describes a seven-dimensional problem that has never been solved in its complete form. 
For neutrino flavor conversions in the early universe one typically assumes initial \emph{space} homogeneity, that allows 
one to reduce the dependence on space-time variables to time only.  
Conversely, for neutrinos in a SN environment, a spatial evolution under the 
assumption of a \emph{stationary} neutrino emission is often considered. For a spherically symmetric neutrino  emission
with negligible variations in the transverse direction, the description further simplifies, the problem being reduced to a
purely radial dynamics. 
However, small space inhomogeneities over the standard rotation and translation invariant background are expected in the early universe, with an initial spectrum in Fourier space very close to the scale invariant Harrison-Zeldovich one, the typical heritage of an inflationary expansion initial stage.
On the other hand, in the SN environment one should account for deviations with respect to a stationary configuration, which are related to hydrodynamical instabilities. 
Both these deviations with respect to the assumed homogeneity conditions can act as seeds for instabilities.

In order to investigate this issue, rather than studying the behavior of  the  complex early universe or SN systems,   we consider here a much more simple toy model, which however  already  illustrates 
the main point of  this paper. Namely,  unless spatial  symmetry (or stationarity) is imposed by hand, the self-interacting neutrino dynamics is unstable with respect to even tiny ripples over a spatially constant (or time independent) background. In particular, we consider a neutrino ensemble in time and one spatial  dimension, initially prepared with equal densities of $\nu_e$ and ${\bar\nu}_e$ evolving in time. In the position invariant case, this system behaves as a flavor pendulum in inverted mass hierarchy.
We then introduce a small space-dependent fluctuation in the matter potential and look for its effect on the evolution of the neutrino density matrix. 
We stress that in our case the matter inhomogeneities just act as a seed to trigger a possible instability and 
are chosen to have amplitudes much smaller than those often considered in literature in relation to the 
MSW effect (see, e.g.,~\cite{Sawyer:1990tw,Kneller:2012id}). 

The paper is organized as follows. In Section II we describe the equations of motion for the neutrino ensemble
evolving in time in  presence of  inhomogeneities. By Fourier transforming the equations in the spatial coordinate, the problem is then reduced to  ordinary differential equations in the time variable for 
the different Fourier modes, which are coupled each other. 
In Section III we numerically solve these equations for a  constant background neutrino potential $\mu > \omega$ and we show how the system exibits a run--away from the flavor pendulum behavior, even for very small matter perturbations. The decoherence is indeed, associated with the growth of  the different Fourier modes that destabilize the ordered pendulum solution. 
We also consider the effect of a time depending neutrino self-potential on this instability. If $\mu$ is a decreasing function, 
as we expect to be the case in both the early universe (with respect to time)
 or SN scenarios (with respect to distance),
the growth of higher wave number Fourier modes might be inhibited, for a sufficiently short decay time scale.
Finally, in Section IV we summarize our results, we comment about the possible effects in realistic neutrino gases and we conclude. 
 
\section{Equations of motion with inhomogeneities}

\subsection{The general formalism}

We start from the equations of motion for a mono-energetic homogeneous  and isotropic two-flavor $(\nu_e, \nu_x)$ relativistic neutrino gas, propagating in time. 
Expanding all quantities in Eq.~(\ref{eq:eom}) in terms of Pauli matrices, one gets the
well-known pendulum equations~\cite{Hannestad:2006nj}
\begin{eqnarray}
\partial_t {\bf P} &=& [+ \omega {\bf B} + \lambda {\bf L} + \mu {\bf D}] \times {\bf P} \,\ , \nonumber \\
\partial_t \bar{\bf P} &=& [-\omega {\bf B} + \lambda {\bf L} + \mu {\bf D}] \times \bar{\bf P} \,\ ,
\label{eq:pend}
\end{eqnarray}
where ${\bf P}$ ($\bar{\bf P}$) are the neutrino (antineutrino) polarization vectors in flavor space. We define as usual ${\bf D} =  {\bf P} - \bar{\bf P}$. 
The vacuum oscillation frequency is $\omega=\Delta m^2/2 E$, 
${\bf L} = e_{\bf z}$, and
$\lambda = \sqrt{2 }G_F n_e$ is the effective potential due to forward scatterings with electrons.  
We remind the reader that a possibly large homogeneous and (in 3 dimensions) isotropic matter term only reduces
the effective mixing angle, and can be rotated away from the equations of motion~\cite{Hannestad:2006nj}.  
The unit vector ${\bf B}$ points  in the mass eigenstate direction in flavor space, such 
that ${\bf B}\cdot{\bf L}=-\cos \theta$, where $\theta$ is the vacuum mixing angle. 
Finally, $\mu \sim \sqrt{2}G_F n_{\nu}$ is the neutrino-neutrino interaction
strength.  

We now consider a non homogeneous background.
In this case the evolution operator in the left-hand-side of Eq.~(\ref{eq:pend}) acquires also a space derivative. In the simplest case of neutrinos propagating in one spatial dimension only, the equation of motion for $\nu$  becomes 
\begin{eqnarray}
(\partial_t  +  \partial_x) {\bf P}(x,t) &=& [+ \omega {\bf B} + \lambda(x,t) {\bf L} \nonumber \\
 &+& \mu (x,t) {\bf D}(x,t)] \times {\bf P}(x,t) ,
\label{eq:evolspace}
\end{eqnarray}
and analogously for $\bar{\bf P}$.  We notice that the third term in the left-hand side of Eq. (\ref{eq:eom}) has been neglected since we are considering a single momentum neutrino ensemble. 
We remark that   in a multi-momentum scenario
it easy to realize that it is of order  $(\lambda_0 p)^{-1}$, with $\lambda_0$ the  length scale over which the background potential in the Hamiltonian is varying, and $p$ is a typical  neutrino momentum. As long as non homogeneities only contain Fourier modes with wavelengths much larger than the neutrino de Broglie wavelength, we have $(\lambda_0 p)^{-1} \ll 1$. Therefore,
this term is expected to be smaller with respect to the other ones in the equations.  

The partial differential equation (\ref{eq:evolspace}) can be transformed into a tower of ordinary differential equations for the Fourier modes
\begin{equation}  
{\bf P}_k (t) =  \int_{-\infty}^{+\infty} dx \,\ {\bf P}(x,t) e^{-i k x} \,\ .
\end{equation}
We find
\begin{eqnarray}
\partial_t {\bf P}_k &=& -i k {\bf P}_k +  \omega {\bf B} \times {\bf  P}_k \nonumber \\
&+&  \int_{-\infty}^{+\infty} \frac{dk^\prime}{2\pi} {\lambda}_{k^\prime}  {\bf L} \times {\bf P}_{k-k^\prime}
 \nonumber \\
&+& \int_{-\infty}^{+\infty} \frac{dk^\prime}{2\pi} 
\int_{-\infty}^{+\infty} \frac{dk^{\prime\prime}}{2\pi} \mu_{k-k^\prime-k^{\prime \prime}} \nonumber \\
& & 
{\bf D}_{k^\prime} \times {\bf P}_{k^{\prime \prime}} \,\ ,
\label{eq:complete}
\end{eqnarray}
where ${\lambda}_{k}(t)$ and ${\mu}_{k}(t)$ are the Fourier transform of ${\lambda} (x,t)$ and ${\mu} (x,t)$, respectively. An analogous set of coupled equations can be written for antineutrino polarization vector  $\bar{\bf P}$.  

\subsection{Monochromatic matter inhomogeneity}

We want to study the simplest model exhibiting an inhomogeneity feature.
We assume that the background neutrino potential $\mu$ is homogeneous, while we parametrize the space 
fluctuations of the matter background by a single wavelength oscillating term of amplitude  
${\epsilon} \ll \mu, \omega$ and wavenumber $k_0$, i.e. 
\begin{equation}
 \lambda=  {\epsilon} \cos (k_{0} x) \,\ .\
\label{eq:matt}
\end{equation}
Of course, this can be generalized to arbitrary
fluctuation smooth profiles, which can be written as a linear superposition of several modes. The Fourier transform  of  Eq.~(\ref{eq:matt}) is quite simple
\begin{equation}
  \lambda_k  =  \epsilon \, \pi \, [\delta(k-k_{0}) + 
\delta(k+k_{0})] \,\ ,
\end {equation}
and Eq. (\ref{eq:complete}) becomes
\begin{eqnarray}
\partial_t {\bf P}_k &=& -i k {\bf P}_k +  \omega {\bf B} \times {\bf  P}_k \nonumber \\
&+& \frac{\epsilon}{2} [{\bf P}_{k-k_{0}}(t) + {\bf P}_{k+k_{0}}(t)] \nonumber \\
&+& \mu \int_{-\infty}^{+\infty} \frac{dk^\prime}{2\pi}  {\bf D}_{k-k'}\times {\bf P}_{k'}  
\label{eq:last1}
\end{eqnarray}
where we have used the fact that since $\mu$ is spatially constant, $\mu_k = 2 \pi \mu \delta(k)$. 

It is easy to see that for a monochromatic 
matter perturbation $\lambda$ as in Eq.~(\ref{eq:matt}), only higher harmonics of the fundamental mode
with $k_n = n k_{0}$ are excited. Defining  $P_n = k_0 P_{k_n}/(2\pi)$, Eq. (\ref{eq:last1}) reduces to a countable tower of coupled equations
\begin{eqnarray}
\partial_t {\bf P}_n &=& -i k_n {\bf P}_n + \omega {\bf B}\times {\bf P}_n \nonumber \\
&+& \frac{\epsilon}{2}  {\bf L}\times [{\bf P}_{n-1} + {\bf P}_{n+1}] \nonumber \\
&+& \mu \sum_{j=-\infty}^{+\infty} {\bf D}_{n-j}\times {\bf P}_j \,\ .
\label{eq:eompert}
\end{eqnarray}
Indeed, we can follow the evolution for  positive modes only, i.e. $n \geq 0$, since the polarization vector ${\bf P}(x,t)$ is a real function and therefore
\begin{equation}
{{\bf P}_n}^{\ast}= {\bf P}_{-n} \,\ .
\end{equation}

To have a clear feeling of how inhomogeneities propagate into the polarization vector, i.e. how $P_n$ and $\bar{P}_n$ for $n\neq 0$ get excited, we explicitly  write  the  two  lowest-order equations for neutrinos, i.e.
\begin{eqnarray}
&&\partial_t {\bf P}_0 =   \omega {\bf B}\times {\bf P}_0 
+ \frac{\epsilon}{2}  {\bf L}\times [{\bf P}_{1}^\ast + {\bf P}_{1}] \nonumber \\
&&+ \mu {\bf D}_{0}\times {\bf P}_0 
+ \mu \sum_{j >0}( {\bf D}_{j}^\ast \times {\bf P}_j+{\bf D}_{j}\times {{\bf P}_j}^\ast ) \,\ , \\
&&\partial_t {\bf P}_1 = -i k_{0} {\bf P}_1 + \omega {\bf B}\times {\bf P}_1 + \frac{\epsilon}{2}  {\bf L}\times [{\bf P}_{0} + {\bf P}_{2}] \nonumber \\
&&+ \mu [{\bf D}_1 \times {\bf P}_0 + {\bf D}_0 \times {\bf P}_1 
+ {\bf D}_2 \times {\bf P}_{1}^\ast + \ldots] \,\ .
\end{eqnarray}
Suppose we start with a perfectly homogeneous initial condition, with only ${\bf P}_0 \neq 0$. The first Fourier mode ${\bf P}_1$ is  then excited, since it is sourced by ${\bf P}_{0}$ in the inhomogeneous matter term. 
All  other modes are then triggered in sequence  in the same way. 
It is also worth noticing that the evolution of the fundamental mode ${\bf P}_0$
is perturbed by the presence of the other Fourier modes ${\bf P}_j$ in its equation.
As we will see, this  leads to a dephasing of the flavor pendulum with respect to the
homogeneous evolution.


\section{Numerical examples}

To illustrate the behavior of self--induced flavor conversions in presence of inhomogeneities we consider a neutrino gas initially composed by 
$\nu_e$ and ${\bar \nu}_e$, only, with  equal densities. As initial condition we take
${P_{0,z}}={\bar P}_{0,z}=1$.
These components represent the space-averaged flavor content in $\nu$ and $\bar\nu$.
All other Fourier modes vanish, ${\bf P}_{n}(0)={\bar{\bf P}}_n(0)=0$ for $n>0$, so the system starts from a spatially constant configuration. 
Numerically, we fix the parameters in Eq.~(\ref{eq:eompert}) at $\mu= 50$, $\omega=1$, $\theta=10^{-2}$ and we consider inverted mass hierarchy  $\Delta m^2 <0$, for which the system is unstable in 
the homogeneous case~\cite{Hannestad:2006nj}. 
In order to get stable numerical results
we follow the evolution 
of the first  $N=20$ Fourier modes. We also solved the equations with larger values of $N$, but the results for the lower modes does not change significantly.
In Fig.~1 we show the flavor evolution 
of the $z$-component of the polarization vector $P_{0,z}$ in the  homogeneous case  with $\epsilon=0$
(dotted curve). In this case, as known, the system oscillates, converting coherently pairs 
$\nu_e \bar\nu_e \leftrightarrow \nu_x \bar\nu_x$ and behaving as a flavor pendulum with
a frequency~\cite{Hannestad:2006nj}
\begin{equation}
\kappa = \sqrt{2 \omega \mu} \,\ .
\end{equation}
In Fig.~1 are also shown the evolution when we switch on an inhomogeneity in the matter potential with $\epsilon = 10^{-7}$ (dashed curve) and
$\epsilon = 10^{-3}$ (continuous curve), respectively.
In this case we have considered as wavenumber of the fluctuation $k_0 = \kappa$.  
Notice that even for a very small inhomogeneity  the coherent behavior of 
the pendulum is broken after some oscillation periods and the system decoheres towards flavor 
equilibrium. 
As expected, increasing the value of the inhomogeneity seed this flavor decoherence is reached earlier.  
 This effect of flavor equilibrium 
is observed in $P_{0,z}$, that  represents the flavor content averaged  over all the space. Indeed $P_z(x,t)$ shows large space fluctuations, due to the interference of various contributions of higher harmonics $k_n$. We expect that considering a more realistic
multi-mode system, with the matter term seed containing several Fourier modes $k_0$, it would easily decohere also in the coordinate space. 

Fig.~2 is a different way to see this phenomenon.  We compare  the evolution of the trajectory of the 
zero mode polarization vector ${\bf P}_0$ in the 
 $x$-$z$ plane for the homogeneous case (left panel) with the case of an inhomogeneous seed with $\epsilon = 10^{-3}$
(right panel).
While in the first case the polarization vector performs 
stable pendular oscillations, keeping its modulus constant, in the inhomogeneous case after few periods, its length shrinks to zero, meaning  that the flavor content averaged over the space coordinate
is equal for two $\nu$ species.  
 
\begin{figure} 
\includegraphics[angle=0,width=1.\columnwidth]{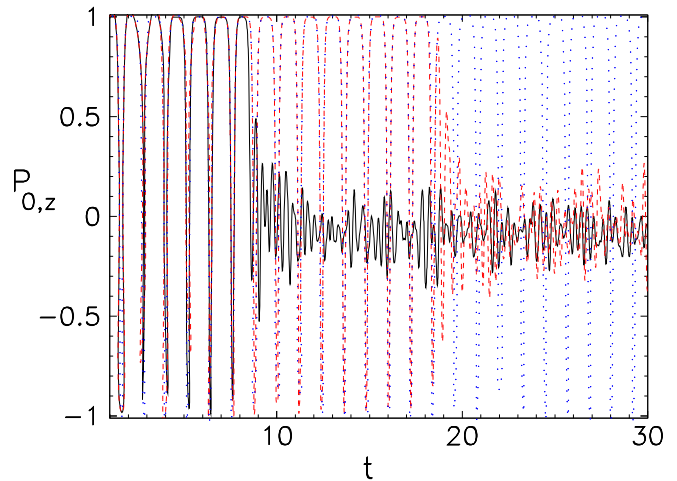}  
\caption{Evolution of the component $P_{0,z}$ for  $k_0= \kappa$.
The continuous curve corresponds to a fluctuation seed $\epsilon=10^{-3}$, 
the dashed one to $\epsilon= 10^{-7}$, while the dotted one is for the 
homogeneous case with $\epsilon=0$.    
 \label{fig1}}
\end{figure}

\begin{figure} 
\includegraphics[angle=0,width=1.\columnwidth]{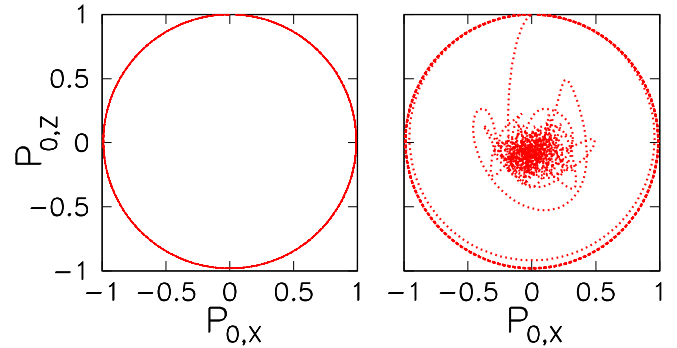}  
\caption{Trajectory of the zero mode polarization vector ${\bf P}_0$ in the $x-z$ plane
 for the homogeneous case (left panel) and for a inhomogeneity
with $\epsilon=10^{-3}$ and $k_0=\kappa$ (right panel).  
 \label{fig2}}
\end{figure}
 
In Fig.~3 we fix $\epsilon =  10^{-3}$ and we rather illustrate 
how changing the  wavenumber of the matter perturbation
affects the onset of the decoherence. With the definition $k_0 = c \kappa$, the continuous  dashed and dotted curves correspond to  $c=1$, $c=10^2$ and $c=10^{-2}$, respectively.  
When the scale of the perturbation $k_0$ is of the order of
the  oscillation scale $\kappa$, the flavor decoherence is approached earlier. Lowering $c$ means considering a longer 
wavelength with respect to the oscillation length scale, and therefore
the neutrino system needs more oscillation cycles to feel the inhomogeneities of the background and eventually decohere. 
On the other hand,  increasing  $c$ the fluctuations tend
to be averaged during an oscillation cycle, and this again tends to shift at larger time the onset of the
decoherence with respect to the case $c=1$.

\begin{figure} 
\includegraphics[angle=0,width=1.\columnwidth]{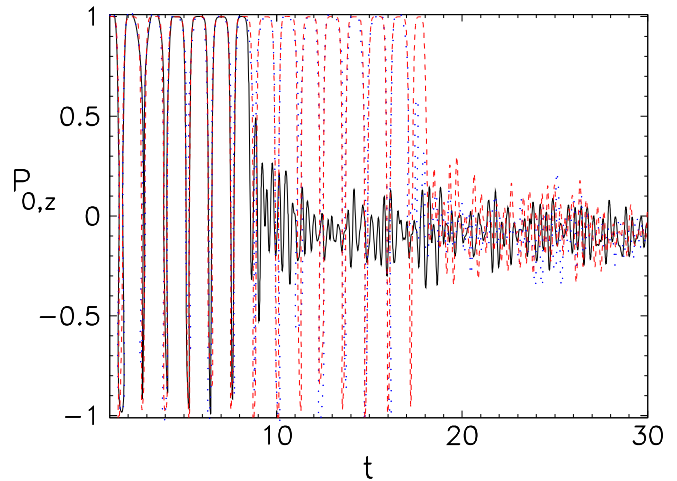}  
\caption{Evolution of the component $P_{0,z}$ for $\epsilon= 10^{-3}$ and different
values of $k_0 = c \kappa$:  $c=1$ (continuous curve),  $c=10^2$
(dashed curve)  and  $c=10^{-2}$ (dotted curve).  
 \label{fig3}}
\end{figure}

\begin{figure} 
\includegraphics[angle=0,width=1.\columnwidth]{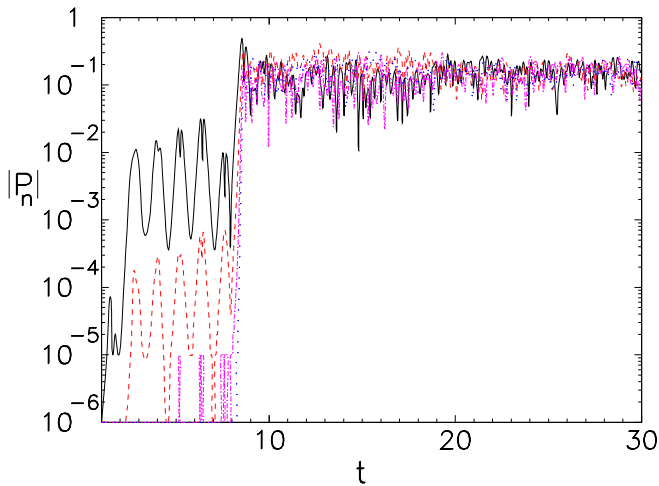}  
\caption{Evolution of the first four modes  $|{\bf P} _n|$ for the case $c=1$
and $\epsilon= 10^{-3}$.  
The continuous curve corresponds to $n=1$, the dashed one to $n=2$, the dash-dotted one to $n=3$ and
the dotted to $n=4$.    
 \label{fig4}}
\end{figure}

The run away of the solution from the stable pendulum behavior is due to the growth of modes with $n>0$, first triggered by the coupling of ${\bf P}_1$ to ${\bf P}_0$ and so on. This is shown in Fig.~4, where we show the  evolution of the modulus of the  first four modes 
$|{\bf P} _n|$ , $n=1,\ldots , 4$   for the case $c=1$.  After ${\bf P}_1$ starts raising,
the higher Fourier modes are also rapidly excited in sequence reaching $|{\bf P} _n| \sim 0.1$.

So far we have assumed that the background neutrino density is independent of time. This guarantees that the non linear term in the kinetic equations keeps always the same order of magnitude, so that eventually all ${\bf P}_n$ are excited, after some sufficiently time laps. In more realistic scenarios, as in the early universe or in   core-collapse SNe, neutrino density is rather expected to be diluted. In the first case, this is due to universe expansion, while for SN it is the effect of the radial matter profile. If $n_\nu$ decays too fast, the system is unable to develop the instability we have discussed so far, and the behavior of the system may be closer to the standard pendulum result. 

To illustrate this point we have considered an exponential decaying term $\mu=\mu_0 \times \exp(-t/\tau)$, with different choices of the characteristic time $\tau$, where we take $\mu_0=50$.
Our results are shown in Fig.~5, having set $\epsilon= 10^{-3}$, and $k_0=\sqrt{2 \mu_0 \omega}$, and with  $\tau=10^3, 10^2, 10$. 
In the left panels we show the evolution of $P_{0,z}$  and the modulus $|{\bf P} _0|$, while
the right panels report the evolution of the first Fourier modes $|{\bf P}_{1}|$ (continuous curve) and $|{\bf P}_{2}|$ (dashed curve).
If the time evolution of $\mu$ is sufficiently slow (i.e. $\tau=10^3$, upper panel) the neutrino ensemble
quickly decoheres, basically as in the case of constant $\mu$. 
For a smaller $\tau=10^2$ (middle panel) the system still shows this behavior, indeed $|{\bf P} _0|$
drops (and the Fourier modes  are excited). However the final
value of $P_{0,z}$ is not zero, but the system tries to follow the slow decay of $\mu$, in a way closer to an homogeneous scenario with $\epsilon=0$. 
Finally, for $\tau=10$ (lower panel) $\mu$ declines too fast to allow the Fourier modes to develop. 
The system does not decohere, $|{\bf P}_0|$ is fixed to unity,  and the behavior is  similar
to what expected for an homogeneous system, i.e. an inversion of the polarization vector with respect
to the initial value. 
Summarizing, we see that decoherence associated with inhomogeneities only grows for a adiabatic evolution of the neutrino background medium, with decay time scales larger than the typical oscillation frequency of the system.

\begin{figure} [!t]
\includegraphics[angle=0,width=1.\columnwidth]{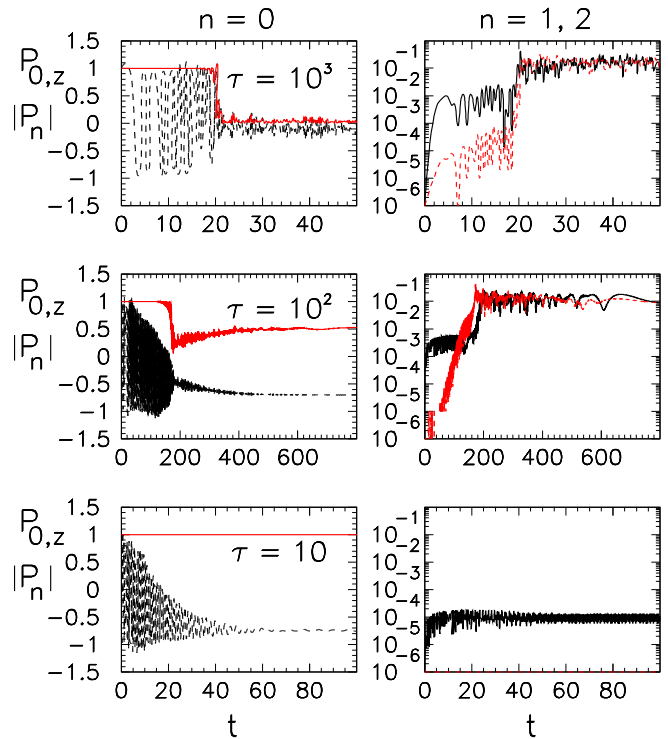}  
\caption{Flavor evolution with a declining neutrino density with 
  $\mu=\mu_0 \times \exp(-t/\tau)$. We take $\mu_0=50$, 
 $\epsilon= 10^{-3}$, and $k_0=\sqrt{2 \mu_0 \omega}$. 
In left panels we show the evolution of $P_{0,z}$ (dashed curves) and 
the modulus $|{\bf P}_0|$ (continuous curves) for different $\tau$, while in 
the right panels we show the evolution of  $|{\bf P}_{1}|$ (continuous curves)
 and $|{\bf P}_{2}|$ (dashed curves).   
 \label{fig5}}
\end{figure}

\section{Conclusions}
The role of symmetries in reducing the complexity of the dynamics of self--interacting neutrino systems has been widely exploited. The general seven--dimensional differential problem can be reduced to more treatable models, and numerically solved with less demanding computation powers. However, simplifying the scenarios to the only time or radial evolution of neutrino density matrix, neglecting space inhomogeneities or non stationary features, provides useful results only if the dynamics is {\it stable} against perturbations. If this is not the case, the behavior of neutrino medium  can be very different from what is found when a particular symmetry is imposed by hand.

In this context we have investigated the emergence of a new kind of instability in the flavor  evolution of a dense neutrino gas, when space homogeneity  is slightly perturbed.
In order to illustrate this effect we have considered the time evolution of a simple system 
 based on an isotropic neutrino ensemble, initially composed by  only $\nu_e$ and
$\bar\nu_e$ with equal densities.
We have  introduced a small amplitude  position-dependent perturbation.
In order to follow the simultaneous temporal and spatial flavor evolution  we have 
Fourier transformed the equations of motion, obtaining a tower of equations for the different modes
associated with the space variable. These modes are coupled because of the effect of the inhomogeneities on 
 the  neutrino-neutrino interaction term. 
We found that in inverted mass hierarchy, where an homogeneous neutrino gas would have evolved according to the  flavor pendulum solution~\cite{Hannestad:2006nj}, the presence of the inhomogeneous term destroys this coherent behavior  and leads to flavor decoherence. This is due to the growing of modes with non zero wavenumber, excited by their coupling to the neutrino homogeneous zero mode. The system is instead  stable in the normal mass hierarchy case.

The instability  discussed here complements the findings of~\cite{decoh}, where for the same $\nu-\bar\nu$
symmetric case, breaking the isotropy of the neutrino propagation leads to a quick decoherence in both mass hierarchies. 
In that case the equations of motion were expanded in multipoles through the Legendre functions and the decoherence
was associated to the excitement of the higher  multipoles. 
Breaking the homogeneity we are observing a similar phenomenology. 

Despite the simplicity of our model -- one space dimension and a monochromatic matter potential disturbance --  we think that features similar to those described in this paper would be also present in more realistic cases. There are indeed, two main frameworks where inhomogeneities (in time or space) could play a role in self-induced flavor conversions:
neutrinos evolution in the early universe or in their streaming off core--collapse supernovae. 
In the first case we know that lepton and neutrino number densities keep the imprint of small space perturbations over the homogeneous background configuration. These are likely produced during the inflationary phase with an almost scale invariant spectrum and initial amplitudes of order $10^{-5}$. In the second example, deviations from a stationary evolution are expected to be triggered by hydrodynamical instabilities.
 
In both these ensembles the neutrino density decreases with respect to the evolution variable. 
In order to mimic this effect,
we have considered a declining matter potential $\mu$. We found that this  could inhibit the growth of the different 
Fourier modes, since the evolution is not enough adiabatic to let them to develop. 
The decoherence effect could be thus, suppressed in realistic scenarios. However, 
it is not guaranteed that  the simultaneous breaking of isotropy and homogeneity may not lead to novel phenomena. 
At this regard
the impact of multi-angle effects  requires a detailed investigation that we leave for a future work. 

Finally, we would like to stress that the formalism we have developed here to treat the simultaneous time and space flavor evolution could be applied to study other deviations from a stationary SN neutrino flavor evolution, 
as those which could be induced by the small backward flux caused by residual neutrino scattering that causes
significant refraction~\cite{Cherry:2012zw,Sarikas:2012vb}. 
In any case it is intriguing that even the simplest neutrino 
flavor pendulum is still a source of new instabilities that were not appreciated before. 
\\
\section*{Acknowledgements} 
A.M. acknowledges G\"{u}nter Sigl for inspiring discussions.
We thank Georg Raffelt for useful comments on the draft.
G.M.   acknowledges support by
the {\it Istituto Nazionale di Fisica Nucleare} I.S. FA51. 
The work of  A.M.   was supported by the German Science Foundation (DFG)
within the Collaborative Research Center 676 ``Particles, Strings and the
Early Universe''.  
N.S. acknowledges support from
the European Union FP7 ITN INVISIBLES (Marie Curie
Actions, PITN- GA-2011- 289442).



\begin{thebibliography}{00}


\bibitem{Duan:2010bg} 
  H.~Duan, G.~M.~Fuller and Y.~-Z.~Qian,
  ``Collective Neutrino Oscillations,''
  Ann.\ Rev.\ Nucl.\ Part.\ Sci.\  {\bf 60}, 569 (2010)
  [arXiv:1001.2799 [hep-ph]].


 \bibitem{Matt}  L.~Wolfenstein,  
				``Neutrino Oscillations In Matter,''  
                Phys.\ Rev.\ D {\bf 17}, 2369 (1978);  
                S. P.~Mikheev and A. Yu.\ Smirnov,  
                ``Resonance Enhancement Of Oscillations In Matter And Solar Neutrino  
				Spectroscopy,''  
                Yad.\ Fiz.\ {\bf 42}, 1441 (1985)  
                [Sov.\ J.\ Nucl.\ Phys.\ {\bf 42}, 913 (1985)].  


\bibitem{Raffelt:2007cb} 
  G.~G.~Raffelt and A.~Y.~.Smirnov,
  ``Self-induced spectral splits in supernova neutrino fluxes,''
  Phys.\ Rev.\ D {\bf 76}, 081301 (2007)
  [Erratum-ibid.\ D {\bf 77}, 029903 (2008)]
  [arXiv:0705.1830 [hep-ph]].

\bibitem{Sawyer:2005jk} 
  R.~F.~Sawyer,
  ``Speed-up of neutrino transformations in a supernova environment,''
  Phys.\ Rev.\ D {\bf 72}, 045003 (2005)
  [hep-ph/0503013].


\bibitem{Banerjee:2011fj} 
  A.~Banerjee, A.~Dighe and G.~Raffelt,
  ``Linearized flavor-stability analysis of dense neutrino streams,''
  Phys.\ Rev.\ D {\bf 84}, 053013 (2011)
  [arXiv:1107.2308 [hep-ph]].

\bibitem{Samuel:1995ri} 
  S.~Samuel,
  ``Bimodal coherence in dense selfinteracting neutrino gases,''
  Phys.\ Rev.\ D {\bf 53}, 5382 (1996)
  [hep-ph/9604341].



\bibitem{Hannestad:2006nj}
  S.~Hannestad, G.~G.~Raffelt, G.~Sigl and Y.~Y.~Y.~Wong,
  ``Self-induced conversion in dense neutrino gases: Pendulum in flavour  
 space,''
  Phys.\ Rev.\  D {\bf 74}, 105010  (2006)
  [Erratum-ibid.\  D {\bf 76},  029901 (2007)]
  [astro-ph/0608695].



\bibitem{Duan:2007fw} 
  H.~Duan, G.~M.~Fuller and Y.~-Z.~Qian,
  ``A Simple Picture for Neutrino Flavor Transformation in Supernovae,''
  Phys.\ Rev.\ D {\bf 76}, 085013 (2007)
  [arXiv:0706.4293 [astro-ph]].


 
  
\bibitem{Qian:1994wh}
  Y.~Z.~Qian and G.~M.~Fuller,
  ``Neutrino-neutrino scattering and matter enhanced neutrino flavor
  transformation in Supernovae,''
  Phys.\ Rev.\  D {\bf 51}, 1479 (1995)
  [astro-ph/9406073].

\bibitem{Duan:2006an} 
  H.~Duan, G.~M.~Fuller, J.~Carlson and Y.~-Z.~Qian,
  ``Simulation of Coherent Non-Linear Neutrino Flavor Transformation in the Supernova Environment. 1. Correlated Neutrino Trajectories,''
  Phys.\ Rev.\ D {\bf 74}, 105014 (2006)
  [astro-ph/0606616].


  
\bibitem{EstebanPretel:2007ec}
  A.~Esteban-Pretel, S.~Pastor, R.~ Tom{\`a}s, G.~G.~Raffelt and G.~Sigl,
  ``Decoherence in supernova neutrino transformations suppressed by
  deleptonization,''
  Phys.\ Rev.\  D {\bf 76}, 125018 (2007)
  [arXiv:0706.2498 [astro-ph]].

\bibitem{Sawyer:2008zs}
  R.~F.~Sawyer,
  ``The multi-angle instability in dense neutrino systems,''
  Phys.\ Rev.\  D {\bf 79} (2009) 105003
  [arXiv:0803.4319 [astro-ph]].
  
  


\bibitem{decoh} 
  G.~G.~Raffelt and G.~Sigl,
  ``Self-induced decoherence in dense neutrino gases,''
  Phys.\ Rev.\ D {\bf 75}, 083002 (2007)
  [hep-ph/0701182].

\bibitem{Raffelt:2013rqa} 
  G.~Raffelt, S.~Sarikas and D.~d.~S.~Seixas,
  ``Axial symmetry breaking in self-induced flavor conversion of supernova neutrino fluxes,''
  Phys.\ Rev.\ Lett.\  {\bf 111}, 091101 (2013)
  [arXiv:1305.7140 [hep-ph]].



\bibitem{Raffelt:2013isa} 
  G.~Raffelt and D.~d.~S.~Seixas,
  ``Neutrino flavor pendulum in both mass hierarchies,''
  Phys.\ Rev.\ D {\bf 88}, 045031 (2013)
  [arXiv:1307.7625 [hep-ph]].


\bibitem{Duan:2013kba} 
  H.~Duan,
  ``Flavor Oscillation Modes In Dense Neutrino Media,''
  Phys.\ Rev.\ D {\bf 88}, 125008 (2013)
  [arXiv:1309.7377 [hep-ph]].


\bibitem{Mirizzi:2013rla} 
  A.~Mirizzi,
  ``Multi-azimuthal-angle effects in self-induced supernova neutrino flavor conversions without axial symmetry,''
  Phys.\ Rev.\ D {\bf 88}, 073004 (2013)
  [arXiv:1308.1402 [hep-ph]].


\bibitem{Mirizzi:2013wda} 
  A.~Mirizzi,
  ``Self-induced spectral splits with multi-azimuthal-angle effects for different supernova neutrino fluxes,''
  arXiv:1308.5255 [hep-ph].


\bibitem{Chakraborty:2014nma} 
  S.~Chakraborty, A.~Mirizzi, N.~Saviano and D.~d.~S.~Seixas,
  ``Suppression of the multi-azimuthal-angle instability in dense neutrino gas during supernova accretion phase,''
  arXiv:1402.1767 [hep-ph].



  
\bibitem{Sigl:1992fn}  
  G.~Sigl and G.~Raffelt,  
  ``General kinetic description of relativistic mixed neutrinos,''  
  Nucl.\ Phys.\ B {\bf 406}, 423 (1993).  
  
\bibitem{Strack:2005ux} 
  P.~Strack and A.~Burrows,
  ``Generalized Boltzmann formalism for oscillating neutrinos,''
  Phys.\ Rev.\ D {\bf 71}, 093004 (2005)
  [hep-ph/0504035].
  
  
\bibitem{Sawyer:1990tw} 
  R.~F.~Sawyer,
  ``Neutrino oscillations in inhomogeneous matter,''
  Phys.\ Rev.\ D {\bf 42}, 3908 (1990).
 
\bibitem{Kneller:2012id} 
  J.~P.~Kneller, G.~C.~McLaughlin and K.~M.~Patton,
  ``Stimulated Neutrino Transformation in Supernovae,''
  J.\ Phys.\ G {\bf 40}, 055002 (2013)
  [arXiv:1202.0776 [hep-ph]].

\bibitem{Cherry:2012zw} 
  J.~F.~Cherry, J.~Carlson, A.~Friedland, G.~M.~Fuller and A.~Vlasenko,
  ``Neutrino scattering and flavor transformation in supernovae,''
  Phys.\ Rev.\ Lett.\  {\bf 108}, 261104 (2012)
  [arXiv:1203.1607 [hep-ph]].
  
\bibitem{Sarikas:2012vb} 
  S.~Sarikas, I.~Tamborra, G.~Raffelt, L.~Hudepohl and H.~-T.~Janka,
  ``Supernova neutrino halo and the suppression of self-induced flavor conversion,''
  Phys.\ Rev.\ D {\bf 85}, 113007 (2012)
  [arXiv:1204.0971 [hep-ph]].

 \end{thebibliography}
\end{document}